\documentclass[aps,prb,preprint,floatfix,showpacs]{revtex4-1}
\usepackage{graphicx,amssymb,amsmath,txfonts}
\usepackage[hang,raggedright]{subfigure}

\newcommand{\Fig}[1]{Fig.~\ref{fig:#1}}

\newcommand{\Sec}[1]{Section~\ref{sec:#1}}
\newcommand{\Eqn}[1]{Eqn.~(\ref{eqn:#1})}
\newlength{\figwidth}
\newcommand*{\MLE}[0]{\theta^\text{MLE}}
\newcommand*{\logL}[0]{\log L}
\DeclareMathOperator{\var}{var}
\DeclareMathOperator{\cov}{cov}

\begin{document}

\title{Database optimization for empirical interatomic potential models}

\author{Pinchao Zhang}
\email{zhang119@illinois.edu}
\author{Dallas R. Trinkle}
\email{dtrinkle@illinois.edu}
\affiliation{Department of Materials Science and Engineering, University of
Illinois, Urbana-Champaign}

\date{\today}
\begin{abstract}
Weighted least squares fitting to a database of quantum mechanical calculations can determine the optimal parameters of empirical potential models. While algorithms exist to provide optimal potential parameters for a given fitting database of structures and their structure property functions, and to estimate prediction errors using Bayesian sampling, defining an optimal \textit{fitting database} based on potential predictions remains elusive. A testing set of structures and their structure property functions provides an empirical measure of potential transferability. Here, we propose an objective function for fitting databases based on testing set errors. The objective function allows the optimization of the weights in a fitting database, the assessment of the inclusion or removal of structures in the fitting database, or the comparison of two different fitting databases. To showcase this technique, we consider an example Lennard-Jones potential for Ti, where modeling multiple complicated crystal structures is difficult for a radial pair potential. The algorithm finds different optimal fitting databases, depending on the objective function of potential prediction error for a testing set. 
\end{abstract}


\maketitle

\section{Introduction}
Atomic-scale simulations have the capability to predict the properties of defect structures that are often inaccessible by experimental techniques.\cite{Daw:1984on,Baskes:1992qt,Xu:1996eh,Daw:1993kx,Wirth:2000qf,Lilleodden:2003ly,Mishin:2001er,Woodward:2008fk} These predictions require accurate and efficient calculations of energies and forces on atoms in arrangements that sample a variety of atomic environments, and may represent even different binding configurations. Accurate quantum mechanical methods are difficult to scale to large systems and long simulation times, while empirical interatomic potentials offer increased computational efficiency at a lower level of accuracy. Maximizing the efficiency of computational material science studies requires the development of potentials that are transferrable, i.e., capable of predicting properties outside their fitting range, and accurate for static and dynamic calculations.

However, without direct transferable derivations of interatomic potentials from quantum mechanical methods, empirical interatomic potentials require high-dimensional non-linear fitting. Many functional forms for empirical potentials have been proposed, including embedded-atom method (EAM)\cite{Daw:1993kx,Li:2003tb}, modified embedded-atom method (MEAM)\cite{Baskes:1989ya,Baskes:1994ys,Baskes:1992qt} and charged-optimized many-body potential (COMB).\cite{Yu:2007ce,Shan:2010zi} There have been multiple implementations of different potential functional forms for various materials.\cite{Mishin:2001er,Lenosky:2000bh,Hennig:2008mb,Park:2012sh,Liu:1996zr,Li:2003tb,Fellinger:2010ys,Sheng:2011vn} Even for the same type of materials, such as Cu\cite{Foiles:1986vy,Zhou:2001iq,Mishin:2001er} and Si\cite{Baskes:1987ws,Baskes:1989ya,Tersoff:1986tf,Balamane:1992qb,Lenosky:2000bh}, different empirical interatomic potential models are proposed for different applications with different transferabilities. There are advanced techniques to optimize the potential parameters based on a weighted least-squares regression to a fitting database of experimental or quantum mechanical calculation data,\cite{Foiles:1986vy,Daw:1993kx} including the force-matching method\cite{Ercolessi:1994vn} for empirical interatomic potential parameter optimization. In force-matching, a fitting database includes quantum mechanical force calculations for diverse atomic environments to obtain realistic empirical potential models. To study the transferability of the empirical potential model, Frederiksen \textit{et al.} applied Bayesian statistics to empirical interatomic potential models: instead of using the best fit, an ensemble of neighboring parameter sets reveal the flexibility of the model.\cite{Frederiksen:2004dg} They showed that the standard deviation of the potential prediction of structure property function is a good estimate of the true error. However, even with these advances, the determination of empirical interatomic potentials relies on the selection and weighting of a fitting database without a clear, quantitative guide for the impact on predictions.

To address the issue of fitting database selection, we present an automated, quantitative fitting-database optimization algorithm based on prediction errors for a testing set using Bayesian statistics. We construct an objective function of the prediction errors in the testing set to optimize the relative weights of a fitting database. This includes the addition or removal of structures to a fitting database when weights change sign.  We demonstrate the viability of the optimization algorithm with a simple interatomic potential model: Lennard-Jones potential fitting of Ti crystal structures. We choose this example as a radial potential has difficulty describing the stability of different crystal structures of a transition metal. The new algorithm also helps to understand the transferability of the empirical potential model for the structures in the testing set.

We start with a brief review of empirical potential models and parameter optimization using a fitting database. Next, we discuss Bayesian error estimation as it applies to our problem. Then we define an objective function with a testing set, and use this quantitative measure to devise an algorithm to optimize a fitting database. Lastly, we demonstrate this new approach on an example system with clear limitations: Lennard-Jones potential for titanium.

\section{Interatomic potential models}
\label{sec:model}

The total energy and forces for the structure of interest are the most basic quantities to calculate since they determine the structural properties. In particular, we are interested in predictions that are derived from energies of atomic arrangements. In atomic-scale simulations, a structure $\alpha$ is a set of atomic positions $\vec{R_m}$ with chemical identities $\chi_m$: $\alpha=\{(\vec{R_m},\chi_m)\}$. The total energy of a structure $\alpha$ is $E_\alpha=E(\{(\vec{R_m},\chi_m)\})$ with forces $\vec{f_\alpha}=-\vec{\nabla}_R E_\alpha$. Density-functional theory (DFT)\cite{Hohenberg:1964zr,Kohn:1965pb} calculations can provide accurate structural energies and forces, but their computational cost limits them to simulation system with at most a few thousand atoms. Other structural properties are derived from energies and forces, and so without loss of generality, we develop our approach based on energies and forces. 

Parameterized empirical interatomic potentials offer a computationally efficient alternative to DFT. Potentials provide approximate energies and forces for atomic configurations that are inaccessibly large for DFT calculations. Generally, an empirical interatomic potential functional can be written as
\begin{multline}
E_\alpha(\theta)\equiv E(\{(\vec{R_i},\chi_i)\};\theta) = \frac{1}{2!}\sum_{mn}V_2^{\chi_m \chi_n}(\vec{R_m}-\vec{R_n};\theta)\\
+\frac{1}{3!}\sum_{mnl}V_3^{\chi_m \chi_n \chi_l}(\vec{R_m}-\vec{R_n},\vec{R_n}-\vec{R_l};\theta)+ \cdots ,
\label{eqn:pot}
\end{multline}
where $\theta$ are parameters, and $V_M^{\chi_1 \ldots \chi_k}$ is an interatomic potential function between $M$ atoms of chemical identity $\{\chi_1, \ldots ,\chi_k\}$. Symmetries of the potential functional form, such as permutation symmetry, translational symmetry, rotational symmetry, etc., can simplify the functional form. A general empirical interatomic potential that reproduces all DFT energy calculations accurately is computational intractable, since it would require a large number of many-body terms. Rather, we are interested in simpler potentials that provide accurate results for a smaller range of atomic configurations including perfect crystals and defect structures under various thermodynamic conditions; this includes potentials that may not be easily written in the form of \Eqn{pot}, such as EAM and MEAM potentials.

The optimal potential parameters $\theta$ derive from comparison to predictions of a database of DFT calculations, and the performance of the potential is evaluated by a testing set of structure property predictions. A fitting database $F$ is a set $\{(\alpha, A_\alpha, w_\alpha)\}$ of structure property functions $A_\alpha$ with an associated structure $\alpha$ and positive (relative) weight $w_\alpha>0$. While a single structure will often have multiple structure property functions with unique weights for fitting, we simplify our notation by indexing on the structure; in what follows, sums over structures $\alpha$ indicate sums over all members of the database $F$. The structure property function $A_\alpha$ may be a scalar such as the energy (relative to a reference structure), vectors such as forces on the atom of the structures, stress tensors and more complicated structure property functions such as lattice constant, bulk modulus or vacancy formation energy or anything that can be defined from the energy $E_\alpha$. In the fitting database, we will compare the structure property functions evaluated using an empirical potential, with the values from DFT, though other choices are possible, such as experimental data. In the weighted least-squares (described later), we impose the trivial constraint $\sum_{\alpha \in F}w_\alpha = 1$, as only relative values of $w_\alpha$ are important. A testing set $T$ is a set $\{(\beta, A_\beta)\}$ of structures $\beta$ with structure property functions $A_\beta$. In a testing set, we will compare structure property functions evaluated using an empirical potential with \textit{either} values from DFT, or using Bayesian sampling of the empirical potential, following Frederiksen \textit{et al.},\cite{Frederiksen:2004dg} which we will discuss in \Sec{bayes}. There are no relative weights for structures in a testing set; rather, these represent a set of predictions whose errors we will evaluate.

In order to assess the prediction errors of the structure property functions, we define the prediction error function $\epsilon_\alpha(\theta)$ as
\begin{equation}
\epsilon^2_\alpha(\theta)=\|A_\alpha(\theta)-A_\alpha\|_2,
\end{equation}
where $A_\alpha(\theta)$ is the structure property function from the empirical atomic potential with parameters $\theta$, $A_\alpha$ is the structure property function from DFT, and $\|\cdot\|_2$ denotes the 2-norm of a $d$-dimensional vector $x$
\begin{equation}
\|x\|_2=\sum_{a=1}^d|x_m|^2.
\end{equation}
We will take the error evaluation of the energy differences between two structures and forces as examples. For energy calculations, the structure property function $A_\alpha$ is
\begin{equation}
A_\alpha=E_\alpha-E_0,
\end{equation}
where $E_0$ is the energy of a reference structure, $0$. The potential energy prediction error is 
\begin{equation}
\epsilon_\alpha(\theta)=\left |(E_{\alpha}(\theta)-E_0(\theta))-(E_{\alpha}-E_0)\right |.
\label{eqn:Energy}
\end{equation}
The force predictions errors are
\begin{equation}
\epsilon^2_\alpha(\theta)=\|f_\alpha(\theta)-f_\alpha\|_2.
\label{eqn:Forces}
\end{equation}
Then, the weighted summed squared error function for a fitting database $F$ is
\begin{equation}
S(\theta, F)=\sum_{\alpha \in F} w_\alpha\epsilon^2_\alpha(\theta),
\end{equation}

\section{Bayesian Error Estimation}
\label{sec:bayes}

We introduce Bayesian sampling to estimate the errors of structure property function predictions and quantitatively analyse the relative weight values in the fitting database. Given a fitting database, we calculate the prediction of structure property function $\langle A_\beta(\theta)\rangle_F$ and the error $\langle \epsilon^2_\beta(\theta)\rangle_F$ of the structure property function. We then derive the analytical expression of the gradient of the Bayesian errors with respect to the weights, $\frac{\partial \langle \epsilon^2_\beta(\theta)\rangle_F}{\partial w_{\alpha}}$. These gradients provide quantitative information on how structure property functions in the fitting database influence the Bayesian predictions of structure property functions in the testing set though weight change. 

Bayesian statistics treats model parameters as random variables with a probability distribution given by a posterior distribution.\cite{Bolstad:2007we} According to the Bayes' theorem, the posterior distribution of the parameters is a product of the prior distribution $\pi(\theta)$ and the likelihood function $L(\theta;F)$,
\begin{equation}
P(\theta;F)\propto \pi(\theta) \times L(\theta;F),
\label{eqn:Bayes}
\end{equation}
where the prior distribution $\pi(\theta)$ includes the information about the potential model before the we take the fitting data into account. Here we use the maximally unbiased prior distribution of a uniform distribution over a measurable set $\mathcal{H}$ of allowed parameters sets,
\begin{equation}
\pi(\theta)=\left[\int_\mathcal{H} d\theta\right]^{-1},
\end{equation}
though other choices are possible. Assuming the errors are independent and identically normally distributed, the likelihood function is\cite{Hogg:2004ce,Frederiksen:2004dg}
\begin{equation}
L(\theta;F) \propto \exp\left(-\frac{1}{W}\sum_{\alpha \in F} w_\alpha\epsilon^2_\alpha(\theta)\right),
\end{equation}
where
\begin{equation}
W=\inf_{\theta} S(\theta, F).
\end{equation}
The log-likelihood function is proportional to the squared error function, $S(\theta, F)$. 
\begin{equation}
\logL(\theta;F)=-\frac{S(\theta, F)}{W}=-\frac{1}{W}\sum_{\alpha \in F} w_\alpha\epsilon^2_\alpha(\theta),
\label{eqn:LogL}
\end{equation}
Since the logarithm is a monotonically increasing function, minimizing $S(\theta, F)$ is equivalent to maximizing the log-likelihood function. The maximum likelihood estimate (MLE) of the parameters $\MLE$ is a function of the fitting database $F$, and $W=S(\MLE, F)$.

The Bayesian prediction of a function $A(\theta)$ is the mean 
\begin{equation}
\langle A(\theta)\rangle_F=\frac{\int P(\theta;F)A(\theta) \, d\theta}{\int P(\theta;F) \, d\theta}=\frac{\int_\mathcal{H} L(\theta;F)A(\theta)d\theta} {\int_\mathcal{H} L(\theta;F)d\theta} .
\label{eqn:mean}
\end{equation}
All the averages are implicit functions of the relative weights in the fitting database. The Bayesian error is the mean squared error of the Bayesian prediction:
\begin{equation}
\langle \epsilon_{A}^2(\theta)\rangle_F=|\langle A(\theta)\rangle_F-A|^2+\var_F(A(\theta)),
\label{eqn:error}
\end{equation}
where $\var_F(A(\theta))=\langle A^2 (\theta)\rangle_F-\langle A(\theta)\rangle_F^2$ is the variance of the Bayesian prediction. The covariance of two functions $A_\alpha(\theta)$ and $A_\beta(\theta)$ represents the correlation between two functions:
\begin{equation}
\cov_F(A_\alpha(\theta),A_\beta(\theta))=\langle A_\alpha(\theta)A_\beta(\theta)\rangle_F-\langle A_\alpha(\theta)\rangle_F\langle A_\beta(\theta)\rangle_F.
\end{equation}
The derivative of a Bayesian prediction with respect to weight is
\begin{equation}
\begin{split}
\frac{\partial \langle A(\theta)\rangle_F}{\partial w_\alpha} &= \frac{\partial}{\partial w_\alpha}\frac{\int_\mathcal{H} L(\theta;F)A(\theta) \, d\theta}{\int_\mathcal{H} L(\theta;F) \, d\theta}\\
&=\cov_F\left(A(\theta), \frac{\partial \logL(\theta;F)}{\partial w_\alpha}\right).
\end{split}
\label{eqn:deriv1}
\end{equation}
Note that
\begin{equation}
\frac{\partial \logL(\theta;F)}{\partial w_\alpha} = - \frac{\epsilon_\alpha^2(\theta) + \frac{\partial W}{\partial w_\alpha}\logL(\theta;F)}{W}.
\label{eqn:deriv2}
\end{equation}
The derivative of $W$ with respect to weight is found using the chain rule,
\begin{equation}
\begin{split}
\frac{\partial W}{\partial w_\alpha}
&=\left.\frac{\partial S(\theta;F )}{\partial w_\alpha}\right|_{\MLE}\\
&\quad +\sum_n \left.\frac{\partial S(\theta;F )}{\partial \theta_n}\right|_{\MLE} \frac{\partial \MLE_n}{\partial w_\alpha}\\
&=\epsilon_\alpha^2(\MLE),
\end{split}
\label{eqn:deriv3}
\end{equation}
as $\MLE$ is an extremum of $S(\theta;F)$. Applying \Eqn{deriv1}--\Eqn{deriv3} to the Bayesian error $\langle \epsilon^2_\beta(\theta)\rangle_F$ yields
\begin{equation}
\begin{split}
\frac{\partial \langle \epsilon^2_\beta(\theta)\rangle_F}{\partial w_{\alpha}}&=\cov_F(\epsilon^2_\beta(\theta), - \frac{\epsilon_{\alpha}^2(\theta) + \epsilon_{\alpha}^2(\MLE)\logL(\theta)}{W}) \\
&=-\frac{1}{W}\left [C^F_{\alpha\beta}-\frac{\epsilon_{\alpha}^2(\MLE)}{W}\cov_F(\epsilon^2_\beta(\theta),S(\theta;F)\right] \\
&=-\frac{1}{W}\left [C^F_{\alpha\beta}-\frac{\epsilon_{\alpha}^2(\MLE)}{W}\sum_\gamma w_\gamma C^F_{\beta\gamma}\right],
\end{split}
\label{eqn:dedw}
\end{equation}
where
\begin{equation}
C^F_{\alpha\beta}=C^F_{\beta\alpha}=\cov_F(\epsilon^2_\alpha(\theta),\epsilon^2_{\beta}(\theta)).
\end{equation}
We define the error of a structure property function in a testing set without DFT calculations by approximating the unknown DFT calculations of the structure property function with its Bayesian prediction. Based on \Eqn{error}, $\langle \epsilon^2(\theta)\rangle_F$ is approximated by $\var_F(A(\theta))$, and so a testing set can include structures \textit{in the absense of} DFT calculations.

We need to evaluate the integral in \Eqn{mean} to calculate the Bayesian predictions and Bayesian error estimation. For complicated high-dimensional, non-linear models such as empirical potentials, the integral cannot be evaluated in closed form, and the high-dimensionality makes direct numerical quadrature converge slowly. We instead use Markov Chain Monte Carlo (MCMC) to numerically integrate. The chain of $N_\text{samples}$ will contain a set of $N$ independent samples $\{\theta_n\}$ (where $N_\text{samples}/N$ is the autocorrelation length), and the numerical estimate of the mean is
\begin{equation}
\langle A_\alpha(\theta)\rangle_F=\frac{\int P(\theta;F)A_\alpha(\theta) \, d\theta}{\int P(\theta;F) \, d\theta}\approx \frac{1}{N}\sum_{n=1}^N A_\alpha(\theta_n),
\label{eqn:MCMC}
\end{equation}
with a sampling error of $\sqrt{\var_F(A_\alpha(\theta))/N}$. Hence, once fitting is complete, the ``best'' set of parameters $\MLE$ defines the empirical potential for predictions, while the ensemble of parameters $\{\theta_n\}$ allows the estimation of errors on those predictions.

\section{Database Optimization}
\label{sec:optimization}

We define an optimal fitting database based on Bayesian errors in the testing set. An empirical potential model should reproduce DFT calculations for a set of atomic environments described by structures in a testing set, and so the Bayesian errors of structure property functions in the testing set are the quantities of interest. Because different types of structure property functions have different units, different error magnitudes, and different degrees of freedom, we need an unbiased choice of objective function to evaluate different fitting database performances based on the Bayesian errors for the same testing set. Here, we consider the difference of the logarithm of the Bayesian errors for one structure property function for two different fitting databases, $F_1$ and $F_2$
\begin{equation}
\ln \langle\epsilon_\beta^2\rangle_{F_1}-\ln\langle\epsilon_\beta^2\rangle_{F_2}=\ln \frac{\langle\epsilon_\beta^2\rangle_{F_1}}{\langle\epsilon_\beta^2\rangle_{F_2}}=\ln\left(1+\frac{\langle\epsilon_\beta^2\rangle_{F_1}-\langle\epsilon_\beta^2\rangle_{F_2}}{\langle\epsilon_\beta^2\rangle_{F_2}}\right).
\end{equation}
If $\langle\epsilon_\beta^2\rangle_{F_1}-\langle\epsilon_\beta^2\rangle_{F_2}$ is small, then $\langle\epsilon_\beta^2\rangle_{F_2}\approx\frac{1}{2}\left(\langle\epsilon_\beta^2\rangle_{F_1}+\langle\epsilon_\beta^2\rangle_{F_2}\right)$, and
\begin{equation}
\ln \langle\epsilon_\beta^2\rangle_{F_1}-\ln\langle\epsilon_\beta^2\rangle_{F_2}\approx \frac{\langle\epsilon_\beta^2\rangle_{F_1}-\langle\epsilon_\beta^2\rangle_{F_2}}{\langle\epsilon_\beta^2\rangle_{F_2}}\approx 2\frac{\langle\epsilon_\beta^2\rangle_{F_1}-\langle\epsilon_\beta^2\rangle_{F_2}}{\langle\epsilon_\beta^2\rangle_{F_1}+\langle\epsilon_\beta^2\rangle_{F_2}}.
\end{equation}
Then the right side of the equation is a relative difference in errors. We propose the objective function of a fitting database $F$ with testing set $T$,  
\begin{equation}
O(F;T)=\sum_{\beta \in T} \ln \langle\epsilon_\beta^2(\theta)\rangle_F,
\label{eqn:Obj_func}
\end{equation}
so that $O(F_1,T)-O(F_2,T)$ is approximately the sum of relative differences in error. Then, the optimal fitting database minimizes the sum log Bayesian errors for a testing set $T$. The objective function is implicitly dependent on the relative weights in the fitting database through the Bayesian error. The gradient of the objective function with respect to weight is analytically calculable (c.f., \Eqn{dedw}). We obtain the optimal weights in the fitting database by minimizing the objective function. Hence we will be able to compare potentials fitted with different fitting databases with respect to the same testing set.

However, the minimum of the objective function can be trivial for pathological fitting databases and testing set combinations. A pathological fitting database and testing set combination is an \textit{underdetermined} fitting database, where the MLE predictions can match the true values of a DFT calculation in both the fitting database and testing set. Thus, if $\langle\epsilon_\beta^2(\theta)\rangle_F\to0$ for any structure $\beta$, then $O(F;T)$ approaches $-\infty$ logarithmically. In order to eliminate the trivial minimum of pathological databases, we introduce a threshold function $t(x)$, 
\begin{equation}
t(x) = \left\{ \begin{array}{ll}
         x & \colon \mbox{$x \geq 2\epsilon^2_0$}\\
        x^2/4\epsilon^2_0+\epsilon^2_0 & \colon \mbox{$x < 2\epsilon^2_0$}\end{array} \right. ,
\end{equation}
that creates a finite minimum of $\langle\epsilon_\beta^2(\theta)\rangle_F$ at $\epsilon_0$. We can choose different error tolerances $\epsilon_0$ for each testing set structure property function. The objective function is then
\begin{equation}
O(F;T)=\sum_{\beta\in T} \ln t(\langle\epsilon_\beta^2(\theta)\rangle_F),
\label{eqn:Obj_func_00}
\end{equation}
and the derivative of the objective function is
\begin{equation}
\frac{\partial O(F;T)}{\partial w_{\alpha}}=\sum_{\beta\in T} \frac{t'(\langle \epsilon_{\beta}^2(\theta)\rangle_F)}{\langle \epsilon_{\beta}^2(\theta)\rangle_F}\frac{\partial \langle \epsilon_{\beta}^2(\theta)\rangle_F}{\partial w_{\alpha}},
\end{equation}
where the derivative calculations are from \Eqn{dedw} and \Eqn{derv_thres}.
The derivative of the threshold function is
\begin{equation}
t'(x) = \left\{ \begin{array}{ll}
         1 & \colon \mbox{$x \geq 2\epsilon^2_0$}\\
        x/2\epsilon^2_0 & \colon \mbox{$x < 2\epsilon^2_0$}\end{array} \right. .
\label{eqn:derv_thres}
\end{equation}
Finally, note that as our likelihood function is independent of $\sum_\alpha w_\alpha$, so
\begin{equation}
\sum_\alpha w_\alpha \frac{\partial O(F; T)}{\partial w_\alpha} = 0.
\end{equation}

The optimal weights are found by minimizing $O(F; T)$, and this includes the addition and removal of structures from the fitting database. According to the definition of the likelihood function, \Eqn{LogL}, the fitting database could include any structures with DFT calculations with a non-negative relative weight value. Structures with positive weight values are structures to fit, and all the other structures that do not contribute to the fitting will have a weight of zero. The optimal weight value can be determined \textit{even for structures not presently in the fitting database}. A structure is added to the fitting database if its optimal weight value is positive, as inclusion of that structure decreases the relative error in the testing set. A structure is removed from the fitting database if its optimal weight value is zero or negative, since removing the structure decreases the relative error in the testing set.

\begin{figure}
\includegraphics[width=\figwidth]{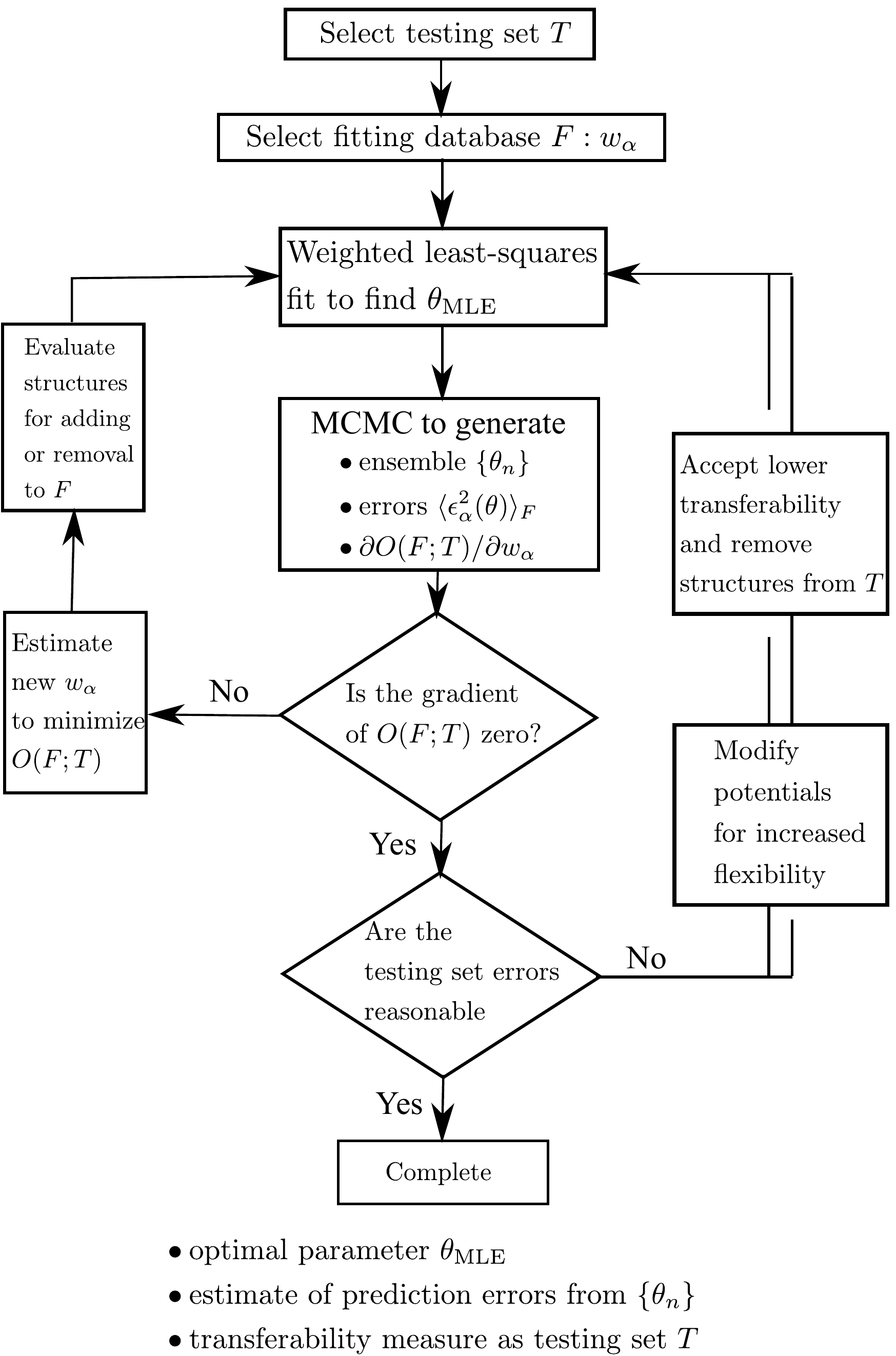}    
\caption{Schematic of a new fitting database optimization algorithm. After selecting a testing set $T$, and an initial fitting database $F$, we find the best set of potential parameters, given by the maximum likelihood estimate $\MLE$. We then use Markov-chain Monte Carlo (MCMC) to generate an ensemble of $\{\theta_n\}$ independent parameters; with this ensemble, we estimate the prediction errors $\langle \epsilon_\beta(\theta)^2\rangle_F$ and compute the gradient of the objective function $O(F; T)$. If the gradients are nonzero, we determine optimal weights $w_\alpha$, as well as consider addition ($w^\text{optimal}_\alpha>0$) or removal ($w^\text{optimal}_\alpha<0$) of structures from the database, and reenter the loop. Once the gradients are zero, we can determine if the testing set errors are acceptable for use; if not, either the range of transferability is lower---indicating a smaller testing set $T$ is needed---or the potential function requires additional flexibility to increase the transferability, and the algorithm is reentered.}
\label{fig:flow_chart}
\end{figure}

\Fig{flow_chart} outlines the new interatomic potential development algorithm based on Bayesian statistics. It starts with a conventional interatomic potential fitting procedure by selecting a potential functional form and DFT structural energies and forces forming a DFT data set. We build the fitting database with a set of structures from the DFT data set and assign each structure with a relative weight. The testing set also contains a set of structures from the DFT data set that test the ability of the potential to model atomic environment outside of the fitting data. We apply non-linear weighted least squares regression to obtain the MLE of parameters of the empirical potential model, and use Markov Chain Monte Carlo (MCMC)\cite{Givens:2005vn} sampling of the posterior distribution to generate the ensemble of parameters around the MLE. We calculate the mean-squared errors of the testing set structures using the parameter ensemble, and construct the objective function and its gradients. Next, we apply a conjugate-gradient method to optimize the objective function and obtain the optimal weights of the fitting database; we can also determine if structures should be added or removed from the fitting database. This step can take advantage of structural searching methods,\cite{Feng:2008} for example, to identify candidate structures, though we do not do so here. We repeat the circuit with the modified relative weight set of the fitting database until the optimal weights converge.

The testing set is the key component of this approach not only because the objective function consists of the mean squared errors of the testing set structures, but also the empirical potential predictions for structures in the testing set should have small errors---whether that is known from comparison with DFT calculations or estimated from Bayesian sampling \textit{without} DFT. With the relative errors in the testing set minimized, any weight deviation from the optimal will result in an increase in relative errors. This means that while we could choose weights to reduce the error of one or several testing set structure property function predictions, it will worsen the predictions of other structures and the trade-off is not worthwhile. Although we are able to optimize the fitting database of the empirical potential models, an optimal fitting database does not guarantee a reliable empirical potential model. The optimization algorithm provides the best possible empirical interatomic potential for a given a fitting database and a given testing set, but it has no judgment on whether the optimal Bayesian errors are acceptable; they can, in fact, be quite large. This can occur if the empirical potential model does not contain the relevant physics to describe the atomic environments in the testing set, which produces reduced transferability. Then, we must---for predictive empirical potential methods---decide to improve the potential model itself to increase transferability or remove structures from the testing set to optimize for reduced transferability.

\section{Implementation on Lennard-Jones Potential fitting of Ti}
\label{sec:LJ-Ti}
\subsection{Potential form and calculation details}
We apply the database optimization algorithm to a simple empirical interatomic potential model, the Lennard-Jones potential. The Lennard-Jones potential is a two-parameter pair potential:
\begin{equation}
\begin{split}
V_2(r;r_0,E_\text{b})=&\left\{ \begin{array}{ll}
         4E_\text{b}\left[\left(\frac{r_0}{r}\right)^{12}-\left(\frac{r_0}{r}\right)^6\right]& \colon \mbox{$r \leq r_{\text{cutoff}}$}\\
       0 & \colon \mbox{$r > r_{\text{cutoff}}$}\end{array} \right.\\
&-V_2(r_\text{cutoff};r_0,E_\text{b})
\end{split}
\label{eqn:LJ}
\end{equation}
where $E_\text{b}$ is the binding energy of a dimer with a separation of $\sqrt[6]{2}r_0$. We choose the cutoff radius $r_{\textrm{cutoff}}=3r_0$, and the allowable parameters are $r_0>0, E_\text{b}>0$.

\begin{table}[b]
\caption{DFT energy calculations of Ti crystal structures. Six different crystal structures were calculated using DFT-GGA. The six crystal structures are hcp, bcc, fcc, simple hexagonal, A15 and $\omega$. The common low temperature phase is hcp, bcc is the high temperature phase, and $\omega$ is a high pressure phase nearly degenerate in energy with hcp.}
\label{tab:Ti_DFT}

\centering

\begin{tabular}{ccccccc}
\hline
\hline
		&hcp &bcc &fcc &hexagonal &A15 &$\omega$\\
\hline
\textit{a} (\AA)  &2.947  &3.261  &4.124  &2.739  &5.192  &4.590\\
\textit{c/a}             &1.583  &N/A  &N/A  &0.999  &N/A  &0.619\\
$E$/atom (eV)	&0.000  &0.108  &0.058  &0.353  &0.192  &--0.005\\
\hline
\end{tabular}
\end{table}

The DFT data set contains six different crystal structures of Ti and the energy versus volume data for all six structures. The DFT calculations are performed with \textsc{vasp}\cite{Kresse:1993wy,Kresse:1996um}, a plane-wave density functional code. We apply a Ti ultrasoft Vanderbilt type pseudopotential,\cite{Vanderbilt:1990}, with a plane-wave cutoff energy of 400eV for energy convergence of 0.3meV/atom.\cite{Hennig:2008mb} The k-point meshes for different structures are, $16\times16\times12$ for hcp, $32\times32\times32$ for bcc, $24\times24\times24$ for fcc, $16\times16\times16$ for hexagonal, $8\times8\times8$ for A15 and $12\times12\times20$ for $\omega$, with Methfessel-Paxton smearing parameter of 0.2eV to obtain an energy accuracy of 1meV/atom.\cite{Hennig:2005kx,Hennig:2008mb} The energy versus volume data includes four different structures with volume of the unit cell as $0.95V_0, 0.975V_0, 1.025V_0$ and $1.05V_0$, where $V_0$ is the unit cell volume of the equilibrium structure. The fitting databases are built from various energy differences and energy versus volume data combinations among the six crystal structures. We generate the Markov chain of the potential parameters using the Metropolis-Hasting algorithm.\cite{Givens:2005vn} The ensemble of the potential parameters contains $10^4$ independent parameter sets from the MCMC simulation with $10^6$ attempted steps, with an auto-correlation length of approximately 100. We use a reweighting scheme discussed in Appendix~\ref{app:reweight} to approximate the objective function values for all possible sets of weights with only one sampling run. Since a radial potential model does not describe the physics of metallic bonding, we expect that the Lennard-Jones potential will not be transferable for testing sets containing many different structures. Our goal is for the algorithm to identify this lack of transferability in the optimization. We systematically consider different types of fitting databases and testing sets with this in mind.

\subsection{Two structured fitting database}
\begin{figure}
\subfigure[]
{\label{fig:binary01}
 \includegraphics[width=0.45\figwidth]{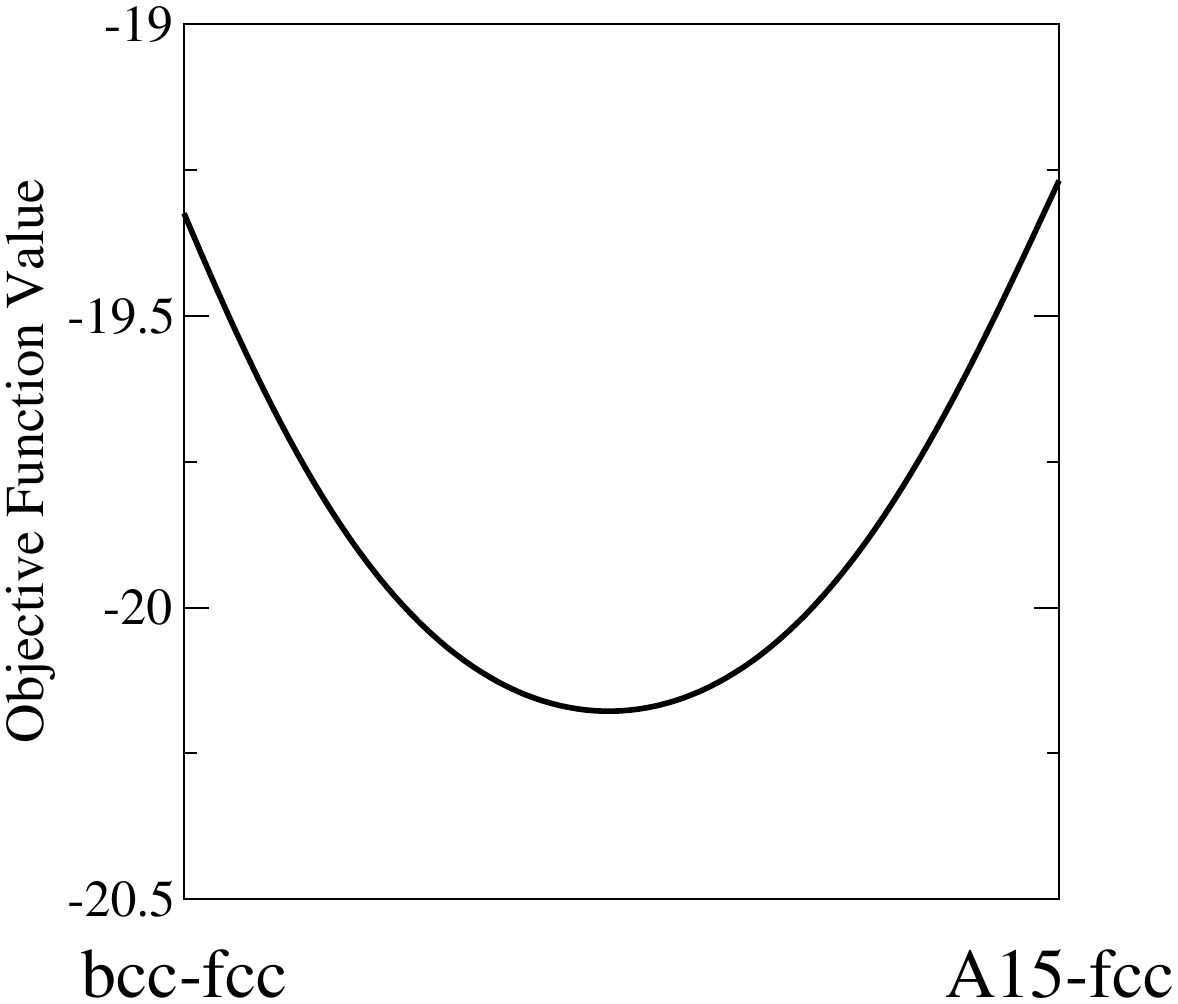}
}
\subfigure[]
{
\label{fig:binary02}
\includegraphics[width=0.45\figwidth]{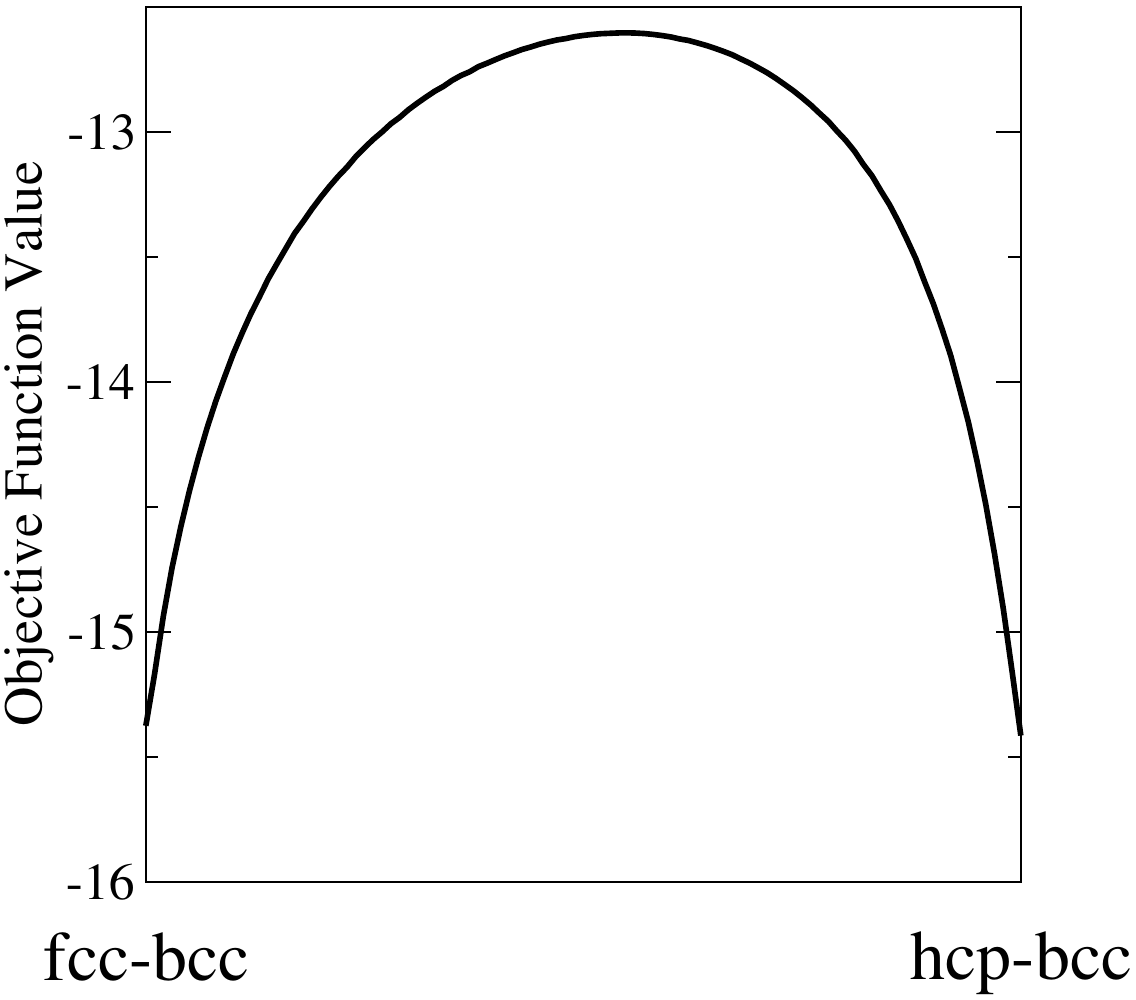}
}
\caption{Two-structured fitting databases of Lennard-Jones potential where the testing set contains the same structures. (a) shows a mixed fitting database with positive optimal weight values, $w_{E_{\text{bcc-fcc}}}=0.485$ and $w_{E_{\text{A15-fcc}}}=0.515$. (b) shows an unmixed fitting database with only one non-zero optimal weight value, which is $E_{\text{fcc-bcc}}$.}
\label{fig:binary}
\end{figure}

We start with a simple fitting database that contains two structural energy differences, and a testing set with the same structures. \Fig{binary} shows two typical objective function behaviors considering all possible weight combinations of two different two-structured fitting database and testing set combinations.  \Fig{binary01} shows the behavior of the objective function of a fitting database with $E_{\text{bcc-fcc}}$ and $E_{\text{A15-fcc}}$. The objective function has a unique minimum with a specific relative weight ratio of the two structures. Moreover, if we calculate the derivative of the objective function with respect to weight at endpoints (where one weight is zero), we can see that each derivative of the objective function with respect to weights indicates that the other structure should be added to the fitting database. Therefore the optimal weight value for both fitting database structures are positive, and we refer this as a ``mixed'' fitting database. On the other hand,  \Fig{binary02} shows a fitting database with $E_{\text{fcc-bcc}}$ and $E_{\text{hcp-bcc}}$. The objective function has minima at the endpoints, which means that a fitting database containing both $E_{\text{hcp-bcc}}$ and $E_{\text{fcc-bcc}}$ has higher relative errors for the testing set than a ``pathological'' fitting databases with only one structure. This is due to the non-transferability between hcp and fcc structures. We refer to these pathological cases as ``unmixed'' fitting databases.

\begin{figure}[b]
\includegraphics[width=\figwidth]{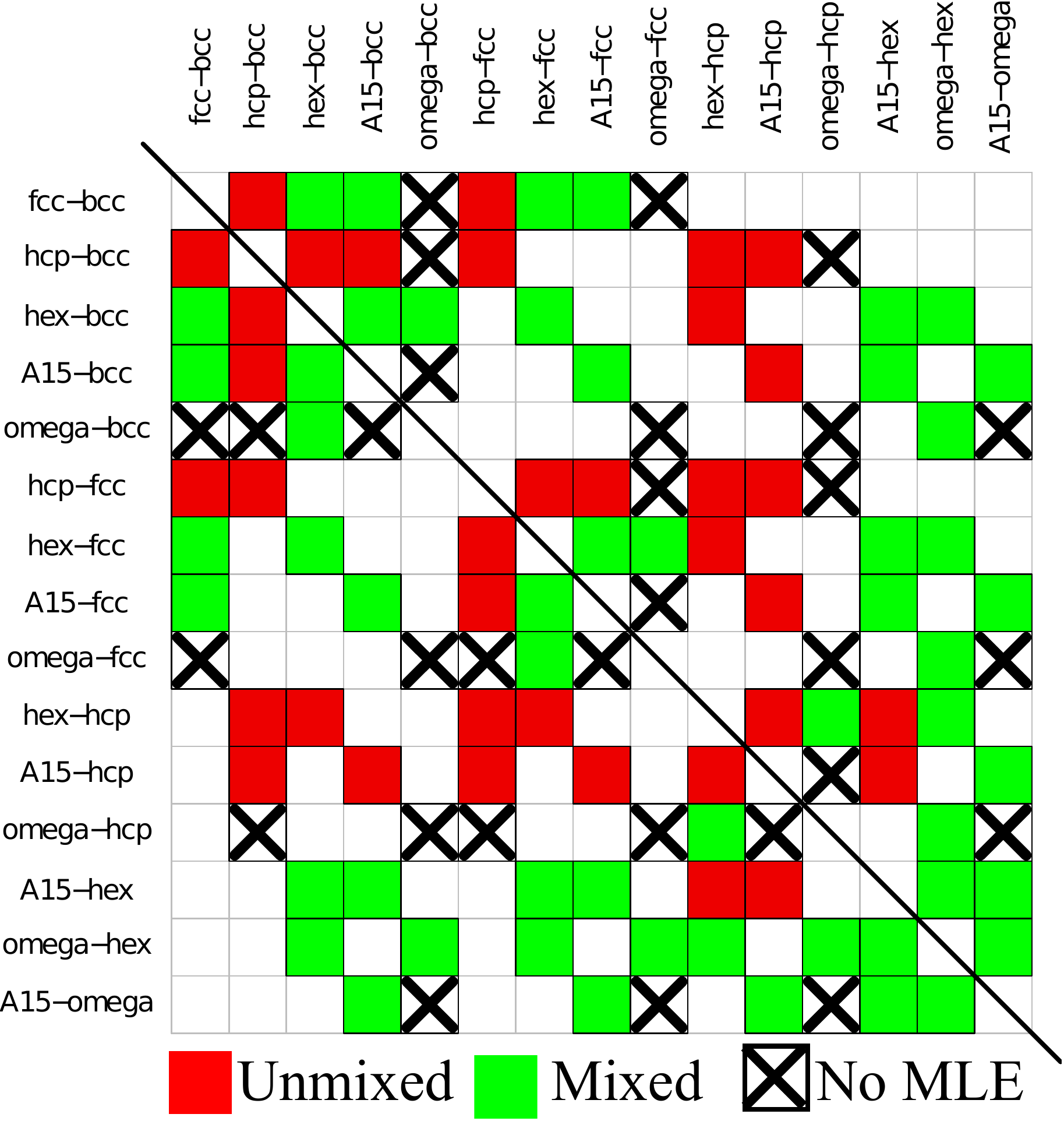}
\caption{All two-structured fitting databases, testing against the same two-structured testing set, where the energy differences share a common reference structure. The green elements represent mixed fitting databases, the red elements indicate unmixed fitting databases, and the $\times$ elements shows databases without a physical MLE.}
\label{fig:binary_all}
\end{figure}

\Fig{binary_all} shows the result of optimizing all possible combinations of two-structured fitting databases with two energy differences sharing a common reference structure. Databases with physical MLEs with positive $E_\text{b}$ and $r_0$ are either mixed or unmixed two-structured fitting databases. Most mixed fitting databases include fcc, bcc, hex and A15 structures and most unmixed fitting databases includes hcp or $\omega$ energy differences. By exploring a wide phase space (six crystal structures of Ti) of Lennard-Jones potential fitting, we have shown that the database optimization algorithm offers an automated, systematic and quantitative way of analyzing empirical potential model fitting with different fitting and testing sets.

\subsection{Three-structured fitting database}
\begin{figure}
\includegraphics[width=\figwidth]{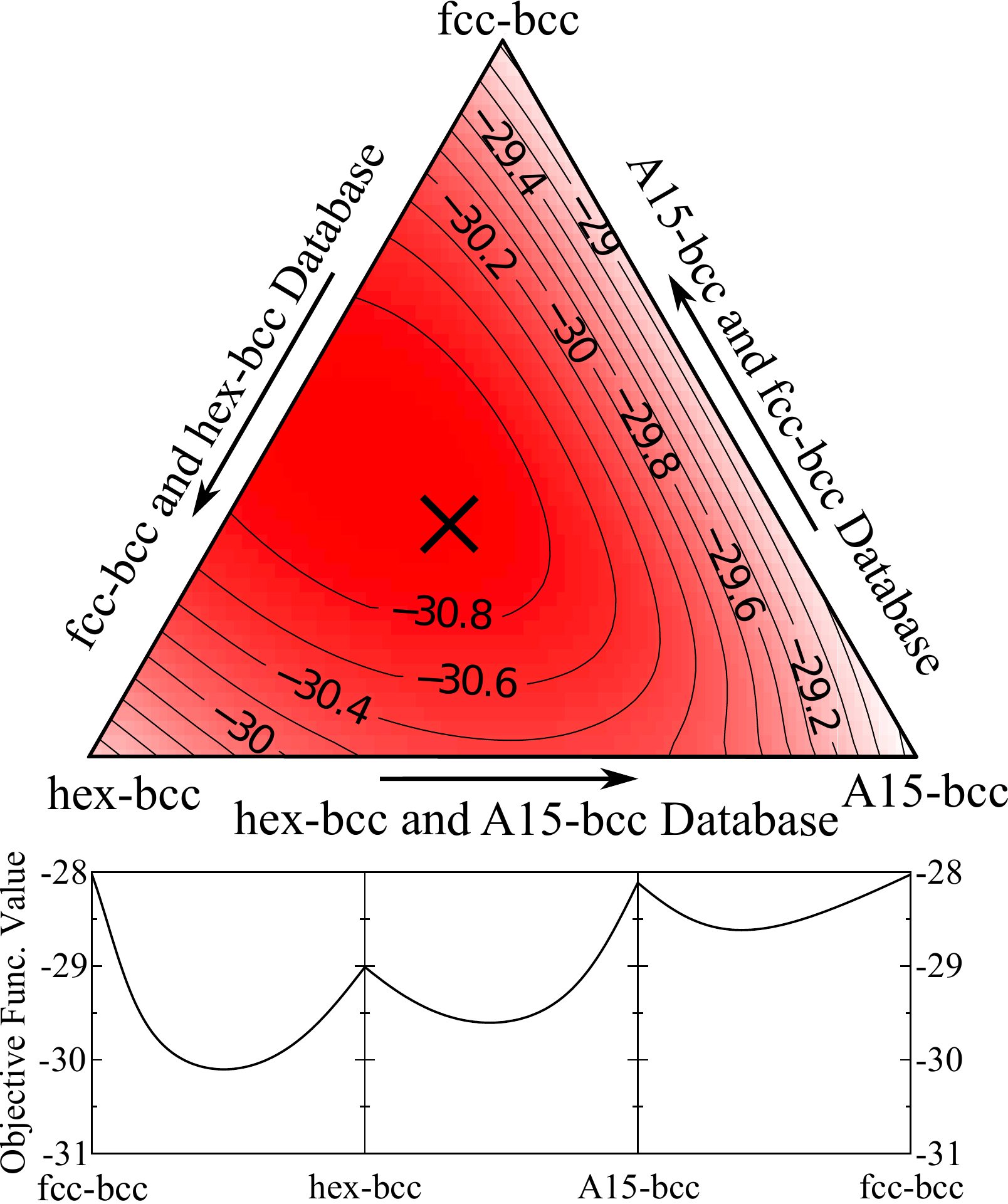}
\caption{Gibbs triangle contour plot of the objective function of the three-structured fitting database fcc+hex+A15 with bcc as reference structure and the testing set contains the same structures. The optimal database, marked by $\times$, is obtained in the interior area with weights $w_{E_{\text{fcc-bcc}}}=0.42$, $w_{E_{\text{hex-bcc}}}=0.42$, $w_{E_{\text{A15-bcc}}}=0.16$.}
\label{fig:fhexA15}
\end{figure}

\begin{figure}
\includegraphics[width=\figwidth]{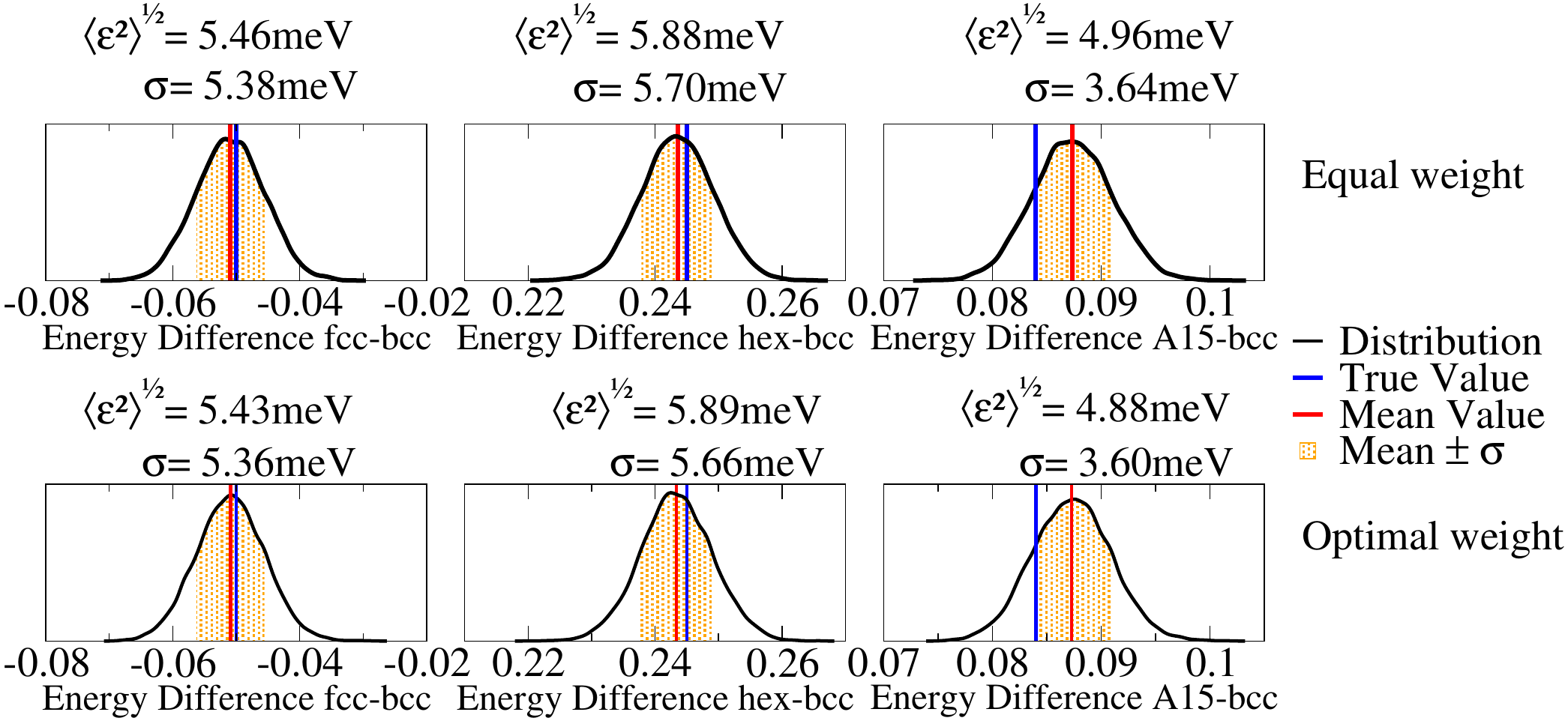}
\caption{Three-structured fitting database prediction distribution for fcc+hex+A15, with bcc as reference structure. The first row of distributions are calculated with equal weights, and the second row are calculated with the optimal weight set, $w_{E_{\text{fcc-bcc}}}=0.42$, $w_{E_{\text{hex-bcc}}}=0.42$, $w_{E_{\text{A15-bcc}}}=0.16$.}
\label{fig:poly01}
\end{figure}

We next apply the database optimization algorithm to the three-structured fitting databases where the testing set contains the same structures as the fitting database. \Fig{fhexA15} is the Gibbs triangle (so that $\sum_\alpha w_\alpha = 1$) contour plot for the objective function for a fitting database constructed from $E_{\text{fcc-bcc}}$, $E_{\text{hex-bcc}}$ and $E_{\text{A15-bcc}}$. All three of the simpler two-structured fitting databases contained in the three-structured fitting database are mixed fitting databases, and there exists a global minimum in the interior of the Gibbs triangle. The optimal weights for all three fitting database structures are positive, with an optimal weight set $w_{E_{\text{fcc-bcc}}}=0.42$, $w_{E_{\text{hex-bcc}}}=0.42$, $w_{E_{\text{A15-bcc}}}=0.16$. If we start with any two-structured fitting database and consider the optimal weight value for the third structure, the gradients of the objective function for the third structure are negative. This indicates that the inclusion of the third structure will decrease the relative errors. \Fig{poly01} shows the comparison of the prediction distributions with equal weights and optimal weights, where the optimal weight set provides reasonable predictions for all three testing set structures. Note, however, that we are only testing energies at \textit{one volume} for each structure; we will consider the addition of volume changes in \Sec{e_vs_vol}.

\begin{figure}
\includegraphics[width=\figwidth]{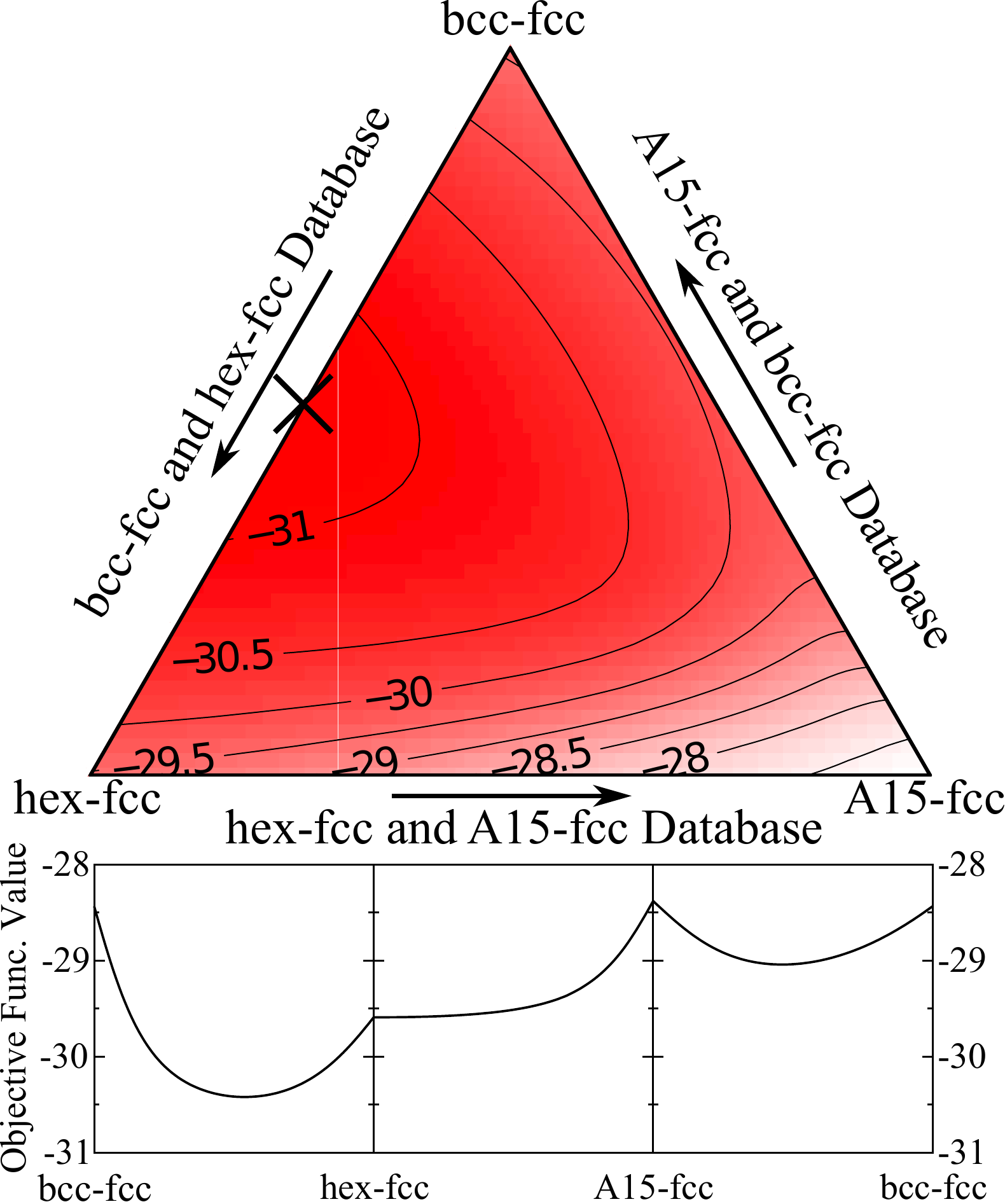}
\caption{Gibbs triangle contour plot of the objective function of the three-structured fitting database bcc+hex+A15, with fcc as reference structure and the testing set contains the same structures. The optimal database, marked by $\times$, is on the edge of $E_{\text{bcc-fcc}}$ and $E_{\text{hex-fcc}}$ for a two-structured fitting database with weights $w_{E_{\text{bcc-fcc}}}=0.46, w_{E_{\text{hex-fcc}}}=0.54$ and $w_{E_{\text{A15-fcc}}}=0$.}
\label{fig:bhexA15}
\end{figure}

\begin{figure}
\includegraphics[width=\figwidth]{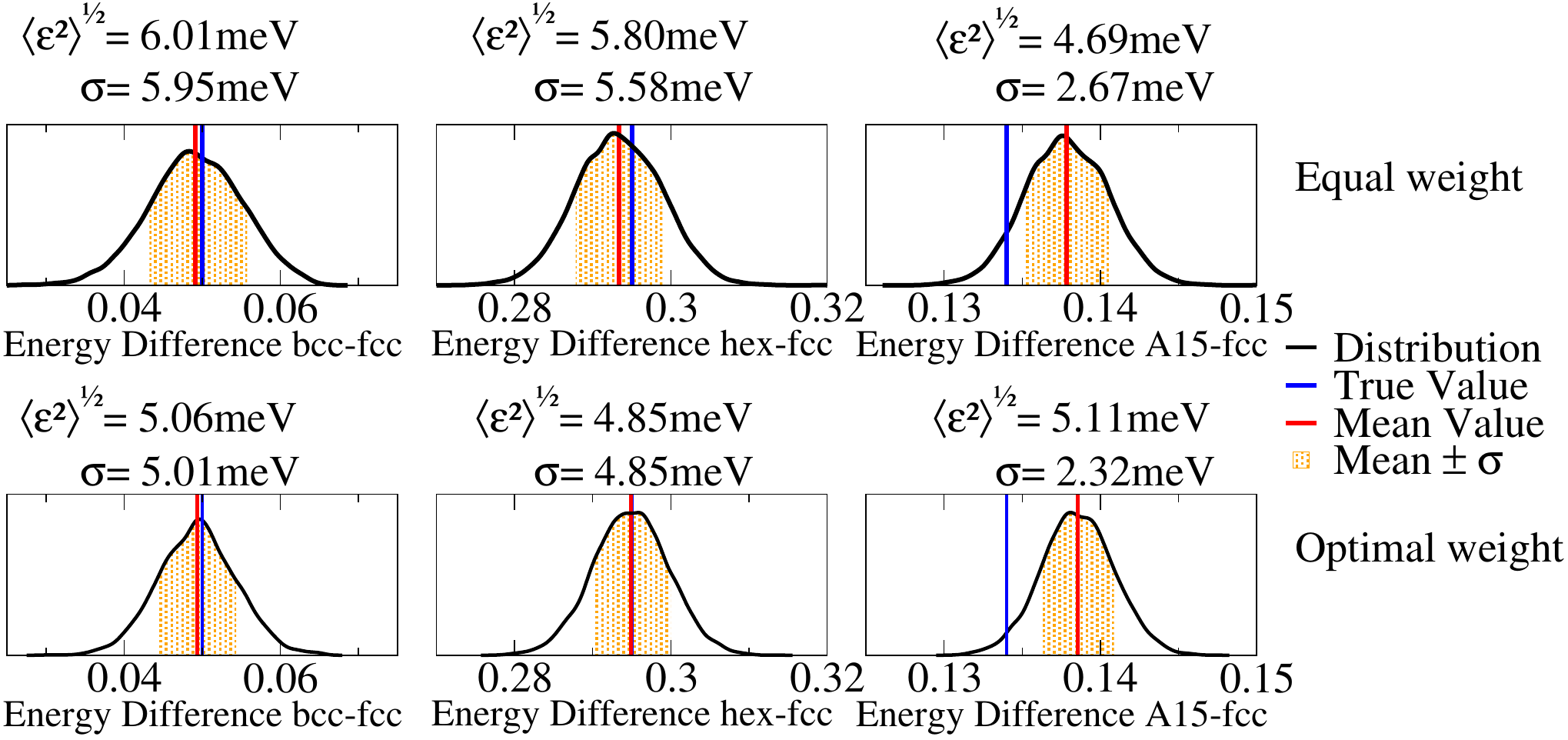}
\caption{Three-structured fitting database prediction distribution for bcc+hex+A15 with fcc as reference structure. The first row of distributions are calculated with equal weights and the second row are calculated with the optimal weight set, $w_{E_{\text{bcc-fcc}}}=0.46, w_{E_{\text{hex-fcc}}}=0.54$ and $w_{E_{\text{A15-fcc}}}=0$.}
\label{fig:poly02}
\end{figure}

\Fig{bhexA15} is a Gibbs triangle for a three-structured fitting database constructed from $E_{\text{bcc-fcc}}$, $E_{\text{hex-fcc}}$, and $E_{\text{A15-fcc}}$. While all three of the two-structured fitting databases are mixed fitting databases, the minimum occurs between $E_{\text{bcc-fcc}}$ and $E_{\text{hex-fcc}}$. The optimal weight values are $w_{E_{\text{bcc-fcc}}}=0.46, w_{E_{\text{hex-fcc}}}=0.54$ and $w_{E_{\text{A15-fcc}}}=0$. The gradient of $w_{E_{\text{A15-fcc}}}$ from the two-structured fitting database $E_{\text{bcc-fcc}}$ and $E_{\text{hex-fcc}}$ is positive meaning fitting to $E_{\text{A15-fcc}}$ will increase the relative errors in the testing set. \Fig{poly02} shows comparison of the prediction distributions evaluated at equal weights and optimal weights. The prediction distribution shows that the optimal testing errors for all three structures are all about 5meV. Therefore, it suggests that for Lennard-Jones potential, one can fit bcc and hex structures to predict A15 structure well.

\begin{figure}
\includegraphics[width=\figwidth]{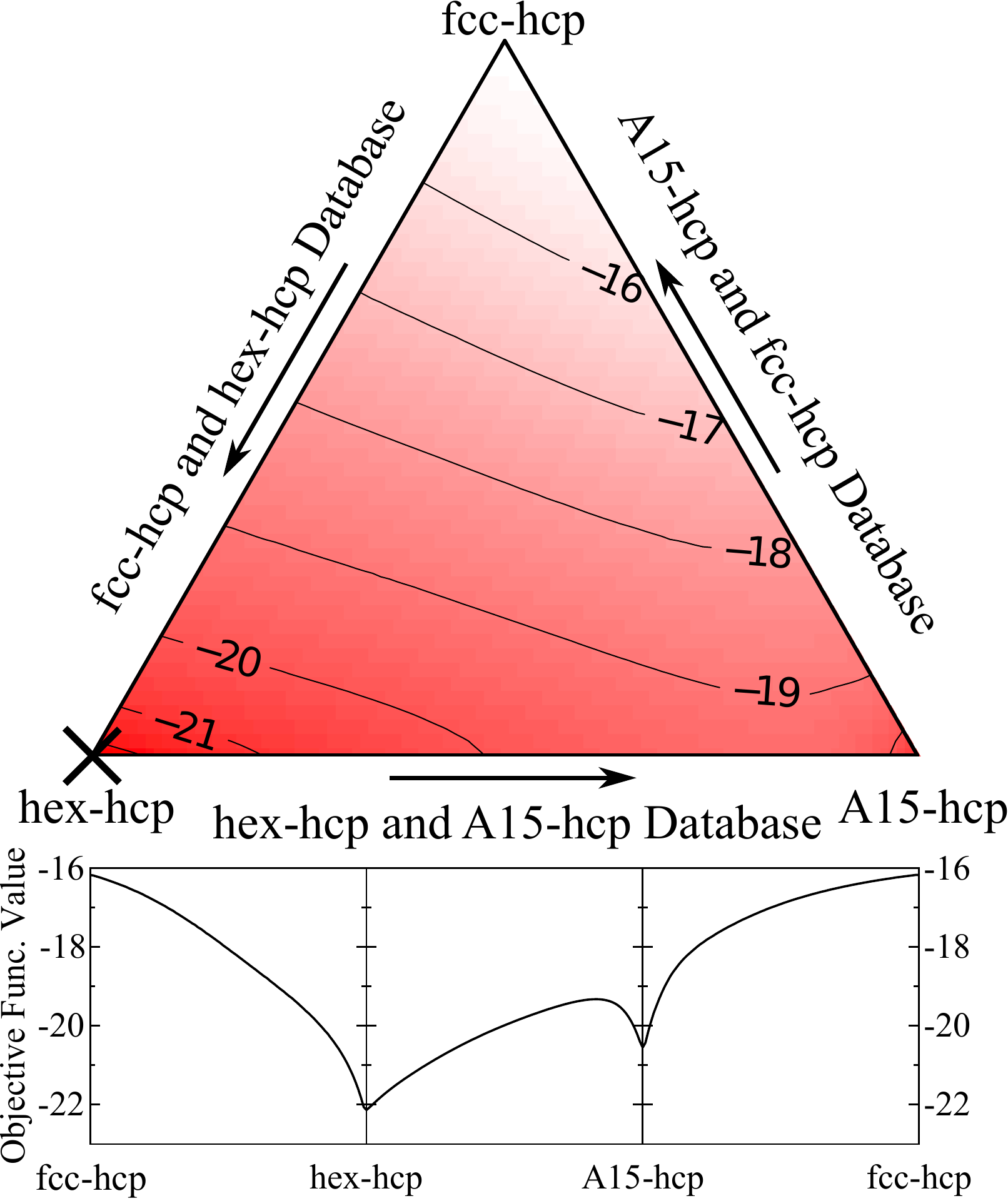}
\caption{Gibbs triangle contour plot of the objective function of the three-structured fitting database, fcc+hex+A15 with hcp as reference structure and the testing set contains the same structures. The optimal database, marked by $\times$, is at the corner of $E_{\text{hex-hcp}}$ for a single database with the optimal weights $w_{E_{\text{bcc-hcp}}}=0, w_{E_{\text{hex-hcp}}}=1$ and $w_{E_{\text{A15-hcp}}}=0$.}
\label{fig:fhexA15hcp}
\end{figure}

\begin{figure}
\includegraphics[width=\figwidth]{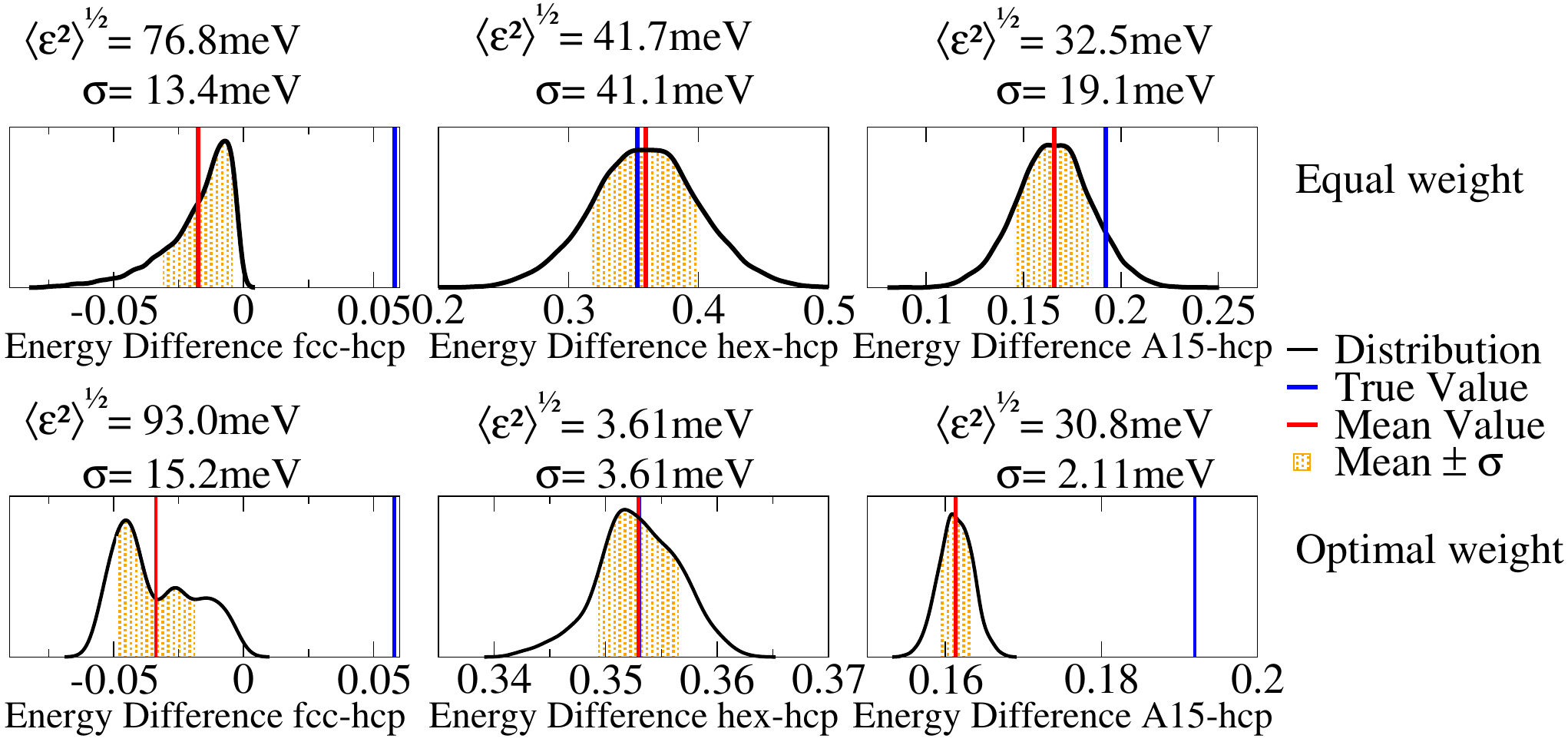}
\caption{Three-structured fitting database prediction distribution for fcc+hex+A15 with hcp as reference structure. The first row of distributions are calculated with equal weights and the second row is calculated with the optimal weight set, $w_{E_{\text{bcc-hcp}}}=0, w_{E_{\text{hex-hcp}}}=1$ and $w_{E_{\text{A15-hcp}}}=0$.}
\label{fig:poly03}
\end{figure}

\Fig{fhexA15hcp} is a Gibbs triangle for a three-structured fitting database constructed from $E_{\text{fcc-hcp}}$, $E_{\text{hex-hcp}}$ and $E_{\text{A15-hcp}}$. Now, all three of the two-structured fitting databases are unmixed fitting databases. Based on the Gibbs triangle contour plot of the objective function, the optimal database includes only one structure $E_{\text{hex-hcp}}$. Adding any of our candidate structures to this will increase the relative errors and worsen the predictions. \Fig{poly03} shows comparison of the prediction distributions evaluated at initial weight and optimal weight. The potential yields a very good prediction for $E_{\text{hex-hcp}}$, but poor estimates for the other two structures. If we use this optimal Lennard-Jones potential to predict $E_{\text{fcc-hcp}}$ and $E_{\text{A15-hcp}}$, the optimal distributions show that the probability of getting the true values are very low. For $E_{\text{fcc-hcp}}$, the true value is 58meV with Bayesian errors of 93meV, and for $E_{\text{A15-hcp}}$, the true value is 192meV with Bayesian error of 30.8meV. It reveals that the optimal Lennard-Jones potential is not transferrable for this testing set, as we expect.

\subsection{Structural energy differences and volume changes}
\label{sec:e_vs_vol}
\begin{figure}
\includegraphics[width=\figwidth]{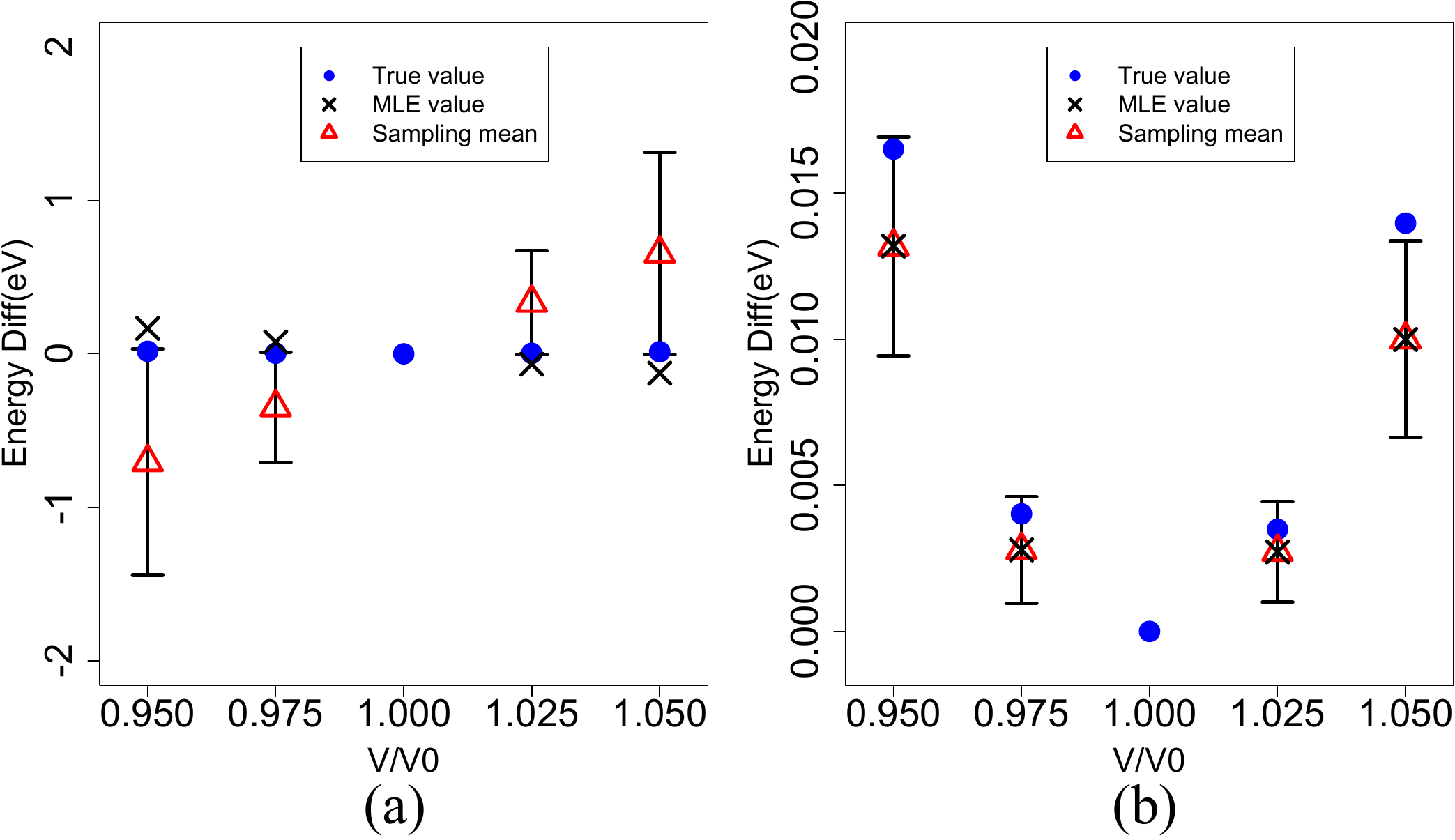}
\caption{Prediction for the hcp energy versus volume curve with fitting database including $E_{\text{fcc-hcp}}$, $E_{\text{hex-hcp}}$, $E_{\text{A15-hcp}}$ and hcp energy versus volume data and testing set including all hcp energy differences and hcp energy versus volume data. $E_\text{hcp}(V)$ with (a) an equal weight set and (b) with the optimal weight set: $w_{E_{\text{hex-hcp}}}=0.548$, $w_{E_\text{hcp}(V)}=0.452$ and all other weights 0.}
\label{fig:e_vol_hcp2}
\end{figure}

Next we apply the algorithm to larger fitting database and testing set combinations. The testing set includes the energy differences between hcp and the other five structures and hcp energy versus volume data. We use four hcp structures with unit-cell volumes of $0.95V_0, 0.975V_0, 1.025V_0$ and $1.05V_0$, for the hcp equilibrium volumes $V_0$. The fitting database starts with the same set of structures and the hcp energy versus volume data. For the four structures representing the hcp energy versus volume data, we constrain their weights to be equal during weight optimization. The $E_{\text{hex-hcp}}$ and hcp energy versus volume data have optimal weight values: $w_{E_{\text{hex-hcp}}}=0.548$ and $w_{\text{hcp-e-vol}}=0.452$, and all other weights are zero. \Fig{e_vol_hcp2} shows that after weight optimization, the inclusion of hcp energy versus volume data improves the prediction of the shape of the energy versus volume curve. It also shows the automatic removal of structures from the database by the optimization algorithm.

\begin{figure}
\includegraphics[width=\figwidth]{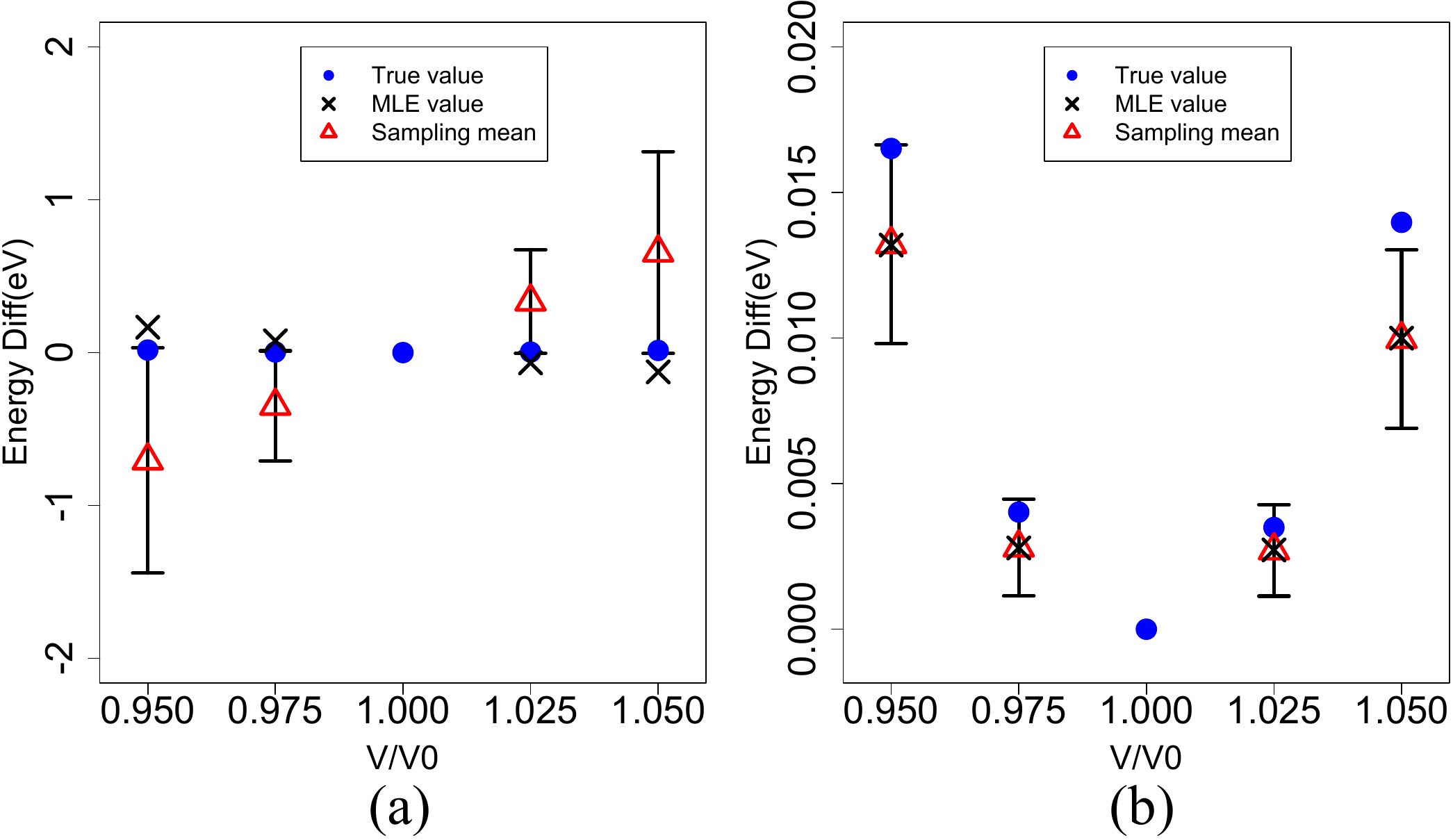}
\caption{Prediction for the hcp energy versus volume curve with fitting database including $E_{\text{fcc-hcp}}$, $E_{\text{hex-hcp}}$, $E_{\text{A15-hcp}}$ and hcp energy versus volume data and testing set including all hcp energy differences and all six energy versus volume data. $E_\text{hcp}(V)$ with (a) an equal weight set and (b) with the optimal weight set: $w_{E_{\text{hex-hcp}}}=0.454$, $w_{\text{hcp-e-vol}}=0.546$ and  and all other weights 0.}
\label{fig:e_vol_hcp3}
\end{figure}

\begin{figure}
\includegraphics[width=\figwidth]{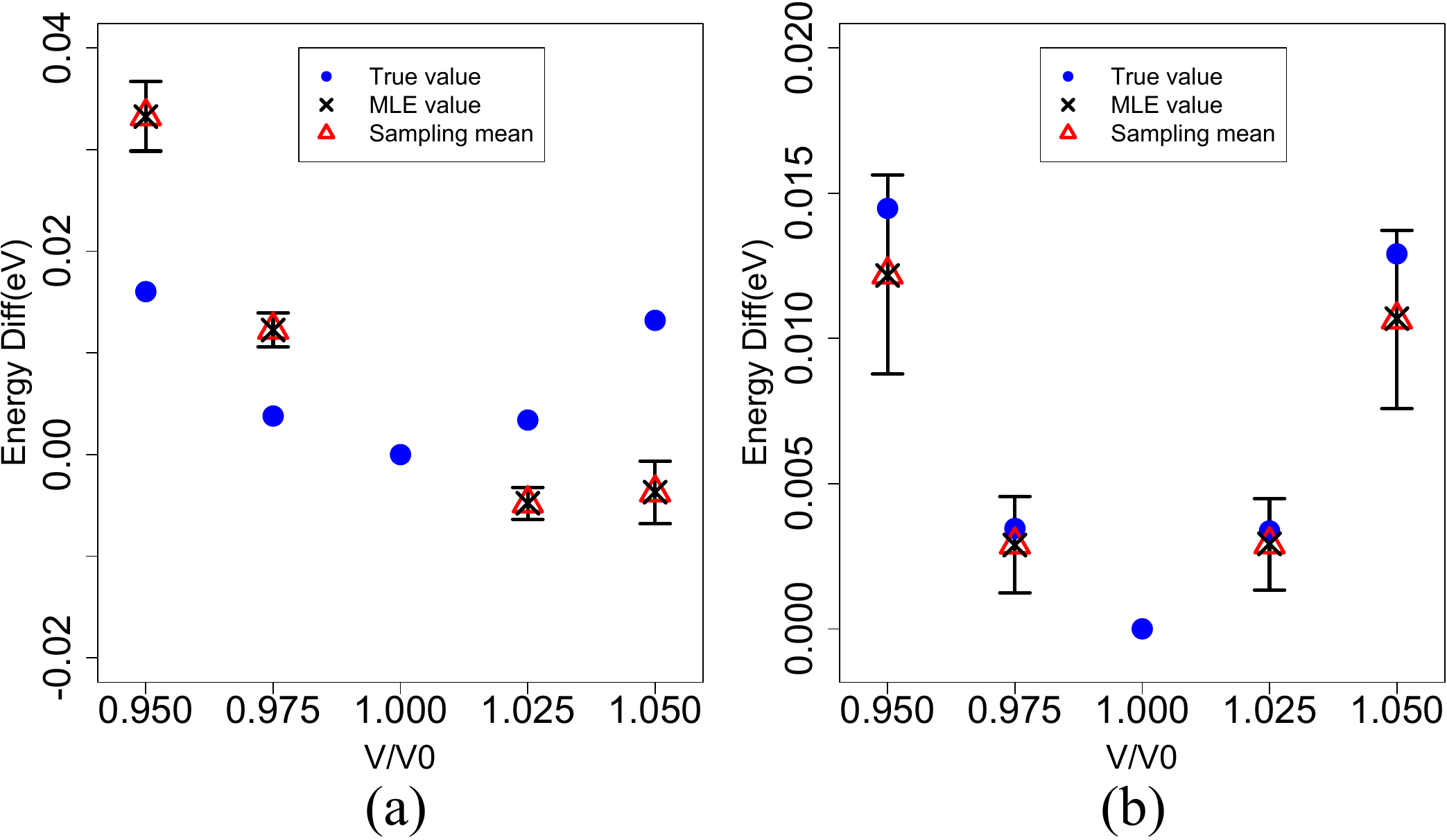}
\caption{Prediction for the bcc and fcc energy versus volume curve with fitting database including $E_{\text{fcc-hcp}}$, $E_{\text{hex-hcp}}$, $E_{\text{A15-hcp}}$, and hcp energy versus volume data, and testing set including all hcp energy differences and all six energy versus volume data. (a) $E_\text{bcc}(V)$ and (b) $E_\text{fcc}(V)$ with the optimal weight set from \Fig{e_vol_hcp3}.}
\label{fig:bf_e_vol}
\end{figure}

\begin{figure}
\includegraphics[width=\figwidth]{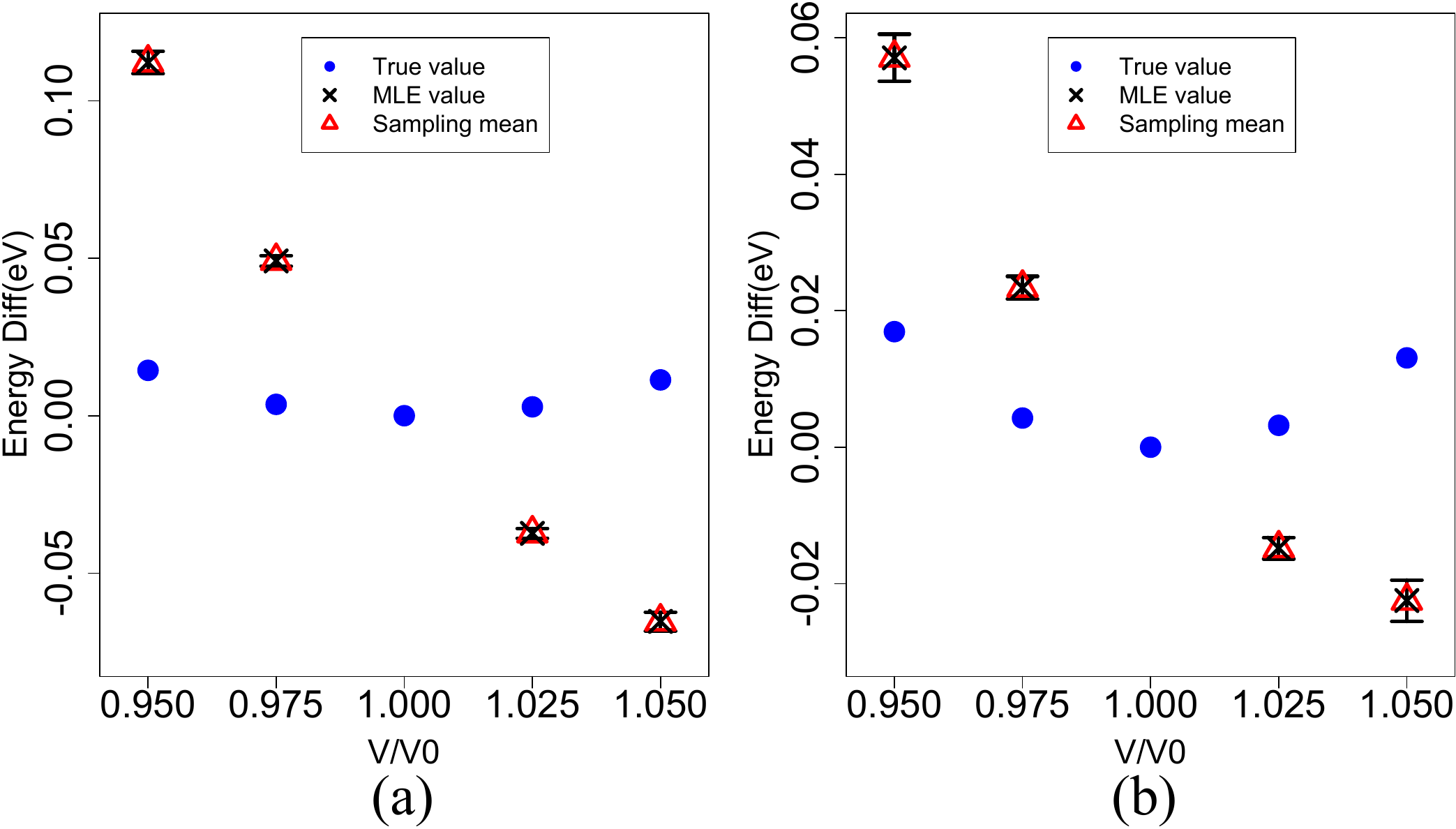}
\caption{Prediction for the hex and $\omega$ energy versus volume curve with fitting database including $E_{\text{fcc-hcp}}$, $E_{\text{hex-hcp}}$, $E_{\text{A15-hcp}}$, and hcp energy versus volume data, and testing set including all hcp energy differences and all six energy versus volume data. (a) $E_\text{hex}(V)$ and (b) $E_\omega(V)$ with the optimal weight set from \Fig{e_vol_hcp3}.}
\label{fig:hexomega_e_vol}
\end{figure}

We next expand our testing set to include energy versus volume data for all six structures along with the structural energy differences. Our fitting database starts with all structural energy differences, and the hcp energy versus volume data---but not the other energy versus volume data. Now, the $E_{\text{hex-hcp}}$ energy difference and hcp energy versus volume data have optimal weight values $w_{E_{\text{hex-hcp}}}=0.454$ and $w_{\text{hcp-e-vol}}=0.546$. In \Fig{e_vol_hcp3}, after weight optimization,  the predictions for hcp energy versus volume data improve significantly compared to the initial equal weight guess. \Fig{bf_e_vol} shows that the optimal fitting database offers a close prediction of the shape of fcc energy versus volume curve, which is expected since the fcc and hcp structures have the same first nearest neighbor atoms. Bayesian error of the four bcc energy differences is too large to have a good predictions for the bcc energy versus volume curve. Similarly, in \Fig{hexomega_e_vol}, sloppy predictions for hex and $\omega$ energy versus volume data are obtained from the optimized fitting database. Hence, the optimal Lennard-Jones potential with the given fitting database and testing set combination does not have enough flexibility to predict energy versus volume curve for bcc, hex and $\omega$.

\subsection{Defect structures without DFT data}

\begin{figure}[b]
\includegraphics[width=\figwidth]{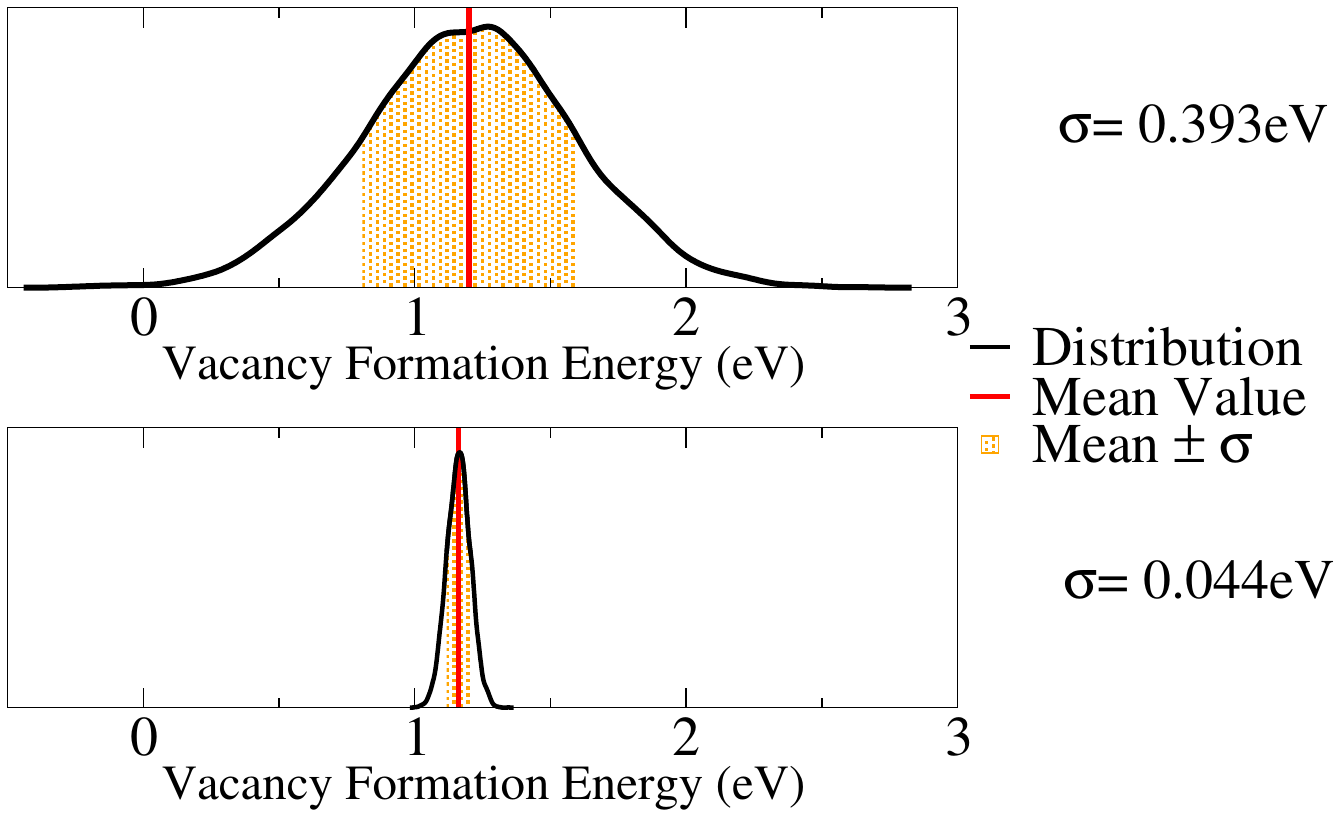}
\caption{Prediction for the hcp vacancy formation energy for two different fitting databases. The fitting database is equally weighted $E_{\text{fcc-hcp}}$, $E_{\text{hex-hcp}}$, $E_{\text{A15-hcp}}$ and hcp energy versus volume data. The bottom is an optimized fitting database where the testing set includes all hcp energy differences, all six energy versus volume data and vacancy configuration. However, the vacancy formation energy from DFT \textit{is not used as input}, and instead is estimated from the potential prediction.}
\label{fig:vacancy}
\end{figure}

Finally, we demonstrate the inclusion of a structure without DFT calculations in the testing set. As we explained in \Sec{bayes}, we assign the structure property prediction as the mean value of the ensemble of the prediction. The ``unknown" structure we add in the testing set is an unrelaxed hcp $4\times4\times3$ supercell containing one vacancy, and the structure property function is the vacancy formation energy. As we are not including DFT data, the comparison is to the mean value of the vacancy formation energy from our empirical potential. The fitting database includes $E_{\text{fcc-hcp}}$, $E_{\text{hex-hcp}}$ and $E_{\text{A15-hcp}}$ and hcp energy versus volume data. The testing set consists of all hcp energy difference, all six energy versus volume data and the single hcp vacancy configuration. \Fig{vacancy} shows the prediction distributions of the vacancy formation energy before optimization and after optimization. The  $E_{\text{hex-hcp}}$ energy difference and hcp energy versus volume data have optimal weight values: $w_{E_{\text{hex-hcp}}}=0.447$ and $w_{\text{hcp-e-vol}}=0.553$. The result shows a significant variance reduction for the vacancy formation energy prediction. It suggests that the prediction will be accurate if the DFT calculation locates within the high likelihood parameter neighborhood of the empirical potential prediction. 

\section{Summary}
\label{sec:summary}

We combine conventional potential parameter optimization methods and the Bayesian sampling technique to propose a new definition of optimal fitting database of empirical interatomic potential models. We choose an objective function as a function of prediction errors for the testing set, and show that minimizing the objective function is equivalent to minimizing the sum of relative errors in the testing set. We optimize the relative weights in the fitting database to minimize the objective function and quantitatively determine the inclusion and removal of structures in the fitting database. Moreover, the performances of two different fitting databases applied on the same testing set can be compared. The algorithm is demonstrated by a simple classical potential model, Lennard-Jones potential fitting of Ti. We go through all possible combinations of two-structured and three-structured fitting databases and analyze the behavior of the objective function with respect to weight change. The new algorithm leads to the best possible empirical interatomic potential model based on the current fitting database and testing set combination. 

\section{Acknowledgements}
The authors thank Yuguo Chen, Henry Wu, and Michael Fellinger for useful comments. This work was supported by the Office of Naval Research through ONR Award No. N000141210752.

\appendix
\section{Reweighting of the sampling chain}
\label{app:reweight}

When weights are changed in the database, we need to reevaluate the mean in \Eqn{mean} using \Eqn{MCMC}. A change to the weights in the fitting database changes the likelihood function to $L(\theta;F^*)$. We rewrite $\langle A(\theta)\rangle_{F^*}$ as
\begin{equation}
\begin{split}
\langle A(\theta)\rangle_{F^*} &=\frac{\int \frac{P(\theta;F^*)}{P(\theta;F)}A(\theta)P(\theta;F) \, d\theta}{\int \frac{P(\theta;F^*)}{P(\theta;F)}P(\theta;F) \, d\theta}=\frac{\int \frac{L(\theta;F^*)}{L(\theta;F)}A(\theta)P(\theta;F) \, d\theta}{\int \frac{L(\theta;F^*)}{L(\theta;F)}P(\theta;F) \, d\theta}\\
&\approx \frac{\sum_{i=1}^N A(\theta_i)\frac{L(\theta_i;F^*)}{L(\theta_i;F)}}{\sum_{i=1}^N \frac{L(\theta_i;F^*)}{L(\theta_i;F)}}.
\end{split}
\end{equation}
Thus a reweighting term is assigned to the original data, and provides new predictions without requiring a new sampling chain.

\end{document}